\documentclass[10pt,conference]{IEEEtran}

\usepackage{listings}
\usepackage{multirow}
\usepackage[]{xcolor}
\usepackage[T1]{fontenc}
\usepackage{graphicx} %
\usepackage{xspace}
\usepackage[nobreak]{mdframed}
\usepackage{pgfmath} %
\usepackage[edges]{forest}
\usetikzlibrary{positioning,arrows.meta,decorations.pathreplacing}
\usepackage{paralist}
\usepackage{hyperref}
\usepackage[english]{babel}
\addto\extrasenglish{
}
\usepackage{pifont}
\usepackage{float}
\usepackage{xurl}
\usepackage{amsmath}
\usepackage{cleveref}%
\usepackage{tablefootnote}
\usepackage{graphicx}
\usepackage{booktabs}
\usepackage{makecell}
\usepackage{color}
\usepackage{tabularray}
\usepackage[htt]{hyphenat}
\usepackage{algorithm}
\usepackage{algpseudocode}
\algnewcommand\algorithmicinput{\textbf{Input:}}
\algnewcommand\Input{\item[\algorithmicinput]}
\algnewcommand\algorithmicoutput{\textbf{Output:}}
\algnewcommand\Output{\item[\algorithmicoutput]}
\usepackage[numbers]{natbib}
\usepackage{balance}
\usepackage{subcaption}
\usepackage{acronym}

\newcommand{\TODO}[1]{\textcolor{red}{#1}\GenericWarning{}{LaTeX Warning: TODO: #1}}\newcommand\todo\TODO

\acrodef{AOT}{Ahead of Time}
\acrodef{JVM}{Java Virtual Machine}
\acrodef{CDS}{Class Data Sharing}
\acrodef{JIT}{Just In Time}

\newcommand{\toolname}{\textsc{LibCache}\xspace}

\definecolor{javared}{rgb}{0.6,0,0} %
\definecolor{javagreen}{rgb}{0.25,0.5,0.35} %
\definecolor{javapurple}{rgb}{0.5,0,0.35} %
\definecolor{javadocblue}{rgb}{0.25,0.35,0.75} %

\lstdefinestyle{customstyle}{
  backgroundcolor=\color{gray!20}, %
  basicstyle=\ttfamily\scriptsize, %
  frame=tb, %
  captionpos=b, %
  breaklines=true, %
  numbers=left, %
  numberstyle=\tiny\color{gray}, %
  commentstyle=\color{black!40}, %
  morecomment=[l]{//}, %
}

\lstdefinestyle{diff}{
    escapechar=\%,
    gobble=4,
}

\begin{document}

\title{Dependencies that Bundle Code and Execution}

\author{
\IEEEauthorblockN{Aman Sharma}
\IEEEauthorblockA{KTH Royal Institute of Technology\\
Stockholm, Sweden\\
amansha@kth.se}
\and
\IEEEauthorblockN{Martin Monperrus}
\IEEEauthorblockA{KTH Royal Institute of Technology\\
Stockholm, Sweden\\
monperrus@kth.se}
}

\maketitle
\pagestyle{plain}

\begin{abstract}
Applications are built from many dependencies.
Each dependency is distributed by package registries but it only contains compiled code and does not ship execution state, so runtimes rebuild that state from scratch on every startup.
Execution caching records this preparation result so subsequent starts can skip it, but no existing system publishes execution state to a registry for downstream consumers to reuse.
We introduce \toolname{}, the first system to distribute per-dependency execution caches through a package registry as first-class artifacts alongside compiled code.
Each library maintainer produces an execution cache from the library's own test suite and publishes it to Maven Central alongside the JAR.
At application build time, \toolname{} merges all per-dependency caches into a single unified cache that covers the classes exercised across every dependency's test suite.
We evaluate \toolname{} on three real-world Java projects across twelve workloads.
\toolname{} speeds up application startup by up to 1.91$\times$ and outperforms a cache built from a single workload in 10 of 12 workloads, with the largest average gain for batik (1.60$\times$ vs.\ 1.27$\times$).
Cache production adds 10--47\% to build time for applications, and execution caches average 18$\times$ the size of the corresponding JAR.
Our main contribution is \toolname{}, which establishes that execution caches are distributable supply chain artifacts and that merging per-dependency caches delivers broader startup coverage than any single observation run.
\end{abstract}

\begin{IEEEkeywords}
Ahead of Time Compilation, Software Supply Chain, Java
\end{IEEEkeywords}

\section{Introduction}

Modern applications are assembled from dozens of third-party dependencies~\cite{soto-valero_multibillion_2022}, each published to a package registry as a compiled code artifact.
When an application starts, the runtime prepares the application and its dependencies in order to execute the code.
A common instance of this concept is shared library loading on a Unix system.
This preparation phase is called ``startup''.
The cost is most painful in short-running workloads, where it can dominate total execution time~\cite{bartoszzaczynski_what_2024,golec_cold_2024}.
This is because the runtime repeats an identical preparation work on every startup~\cite{mehta_reusing_2023}.
Execution caching is a solution to this, it consists of recording precomputed startup artifacts that subsequent starts skip the cached steps and are faster~\cite{lion_dont_2016}. Pycache is an example of such a technology in the context of Python.

Existing execution caches, however, only cover the code paths exercised during a single observation run, done to create the cache~\cite{lam_jep_2025a}.
Any workload that exercises code paths absent from that run pays the full preparation cost for those paths.
This problem compounds with the application's dependency footprint, as each dependency independently exposes multiple execution paths that a single observation run cannot capture exhaustively.
Execution caching, as done today, is remarkably behind our modern software paradigm where applications are made of many dependencies.
We need to rethink execution caching in a world with dependencies.
We need to change the way we build, distribute, and start software packages.

In this paper, we claim that dependencies should carry more than code only.
They should bundle code with execution data.
We introduce \toolname{}, the first system to realize this vision end to end.
Each library maintainer produces an execution cache from the library's own test suite and publishes it to the package registry alongside the compiled library.
When an application is built by a downstream user, the package manager downloads each dependency with its execution cache, and \toolname{} combines all per-dependency caches into a single unified execution cache.
Finally, the application starts with this merged cache.
The \toolname{} cache spans the classes exercised across every dependency's test suite, offering broad coverage.

We evaluate \toolname{} on three real-world Java applications (batik, thymeleaf, biojava), each with four distinct workloads.
\toolname{} achieves up to 1.91$\times$ startup speedup over the no-cache baseline and outperforms a single-workload AOTCache in 10 of 12 workloads, with the largest average gain for batik (1.60$\times$ vs.\ 1.27$\times$), though for two lightweight biojava workloads the difference is within measurement noise.
Cache production adds between 10\% and 47\% to the build time of the main application module; dependencies vary more widely depending on test suite size and presence.
Execution caches average 18$\times$ the size of the corresponding JAR.
In other words, our prototype implementation \toolname{} fully demonstrates that our vision of bundling code with execution is feasible and effective, end to end.

No prior system distributes independently produced, per-dependency execution caches through a package registry, enabling merging of execution caches at build time.
The closest related work is JITServer~\cite{khrabrov_jitserver_2022}, which shares JIT-compiled code across multiple JVM instances.
Unlike \toolname{}, JITServer compiles code reactively as it executes, so a fresh deployment has no coverage until the system accumulates sufficient runtime observations.

Our contributions are the following.

\begin{itemize}
  \item \toolname{}, the first system to distribute per-dependency execution caches through Maven Central, enabling composable cache merging at application build time.
  \item A merge algorithm that combines per-dependency execution caches into a single unified project cache.
  \item An experimental design for evaluating execution caches across diverse workloads, applied to three real-world Java projects.
  \item An open-source implementation as a fork of OpenJDK with a publicly available replication package.
\end{itemize}

\section{Background}

\subsection{JVM Start-Up Workflow}

Java compiles source code to platform-independent bytecode, which the \ac{JVM} executes at runtime.
When starting a Java application, the \ac{JVM} performs four steps.

\textbf{Loading} searches the classpath for the \texttt{.class} file which contains the \ac{JVM} bytecode.
Then this bytecode is parsed into an internal \ac{JVM} metadata structure and allocated in memory.

\textbf{Linking} takes this metadata structure through three steps~\cite{lindholm_chapter_2025}.
First, the \ac{JVM} verifies bytecode correctness.
Second, it prepares the class by allocating static fields and zeroing them to their type's default value (e.g., \texttt{0} for integers, \texttt{null} for references).
Third, it resolves symbolic references in the constant pool to concrete memory addresses.

\textbf{Initialization} executes the class's static initializer (\texttt{<clinit>}), which assigns programmer-specified values to static fields and runs static initialization blocks, populating heap objects.

\textbf{\ac{JIT} Compilation} begins once the application is running.
The \ac{JVM} first interprets bytecode while collecting profiling data, such as method invocation counts.
It then compiles frequently-invoked methods using \ac{JIT} compilers.
Because the \ac{JVM} must first observe execution before compiling, the application runs at interpreted speed until the \ac{JIT} produces optimized native code.
This period is known as \ac{JVM} \emph{warmup}.

\subsection{Execution Caching}
\label{sec:execution-caches}

Caching is a fundamental technique in computer systems that consists of storing the result of a computation so that it can be reused later, trading storage space for reduced execution time.
It is ubiquitous across the software stack, spanning CPU caches that reduce memory access latency~\cite{smith_cache_1982} and storage caches that avoid repeated disk I/O~\cite{zhang_distributed_2012}, all the way to application-level systems such as database query caches, web browser caches~\cite{mozilla_http_2026}, and developer toolchains~\cite{gradle_build_2026, mattvollmer_guide_2025}.
One important target for caching is startup in language runtimes, because the runtime repeats many identical steps on every startup~\cite{wimmer_initialize_2019,xu_sharejit_2018}.
In the context of runtime startups, an \emph{execution cache} stores the result of an initial recorded run so that subsequent startups skip the cached steps.

Table~\ref{tab:execution-caches} surveys execution caches across popular language runtimes.
For each runtime, the table records the startup stage eliminated by the cache, the package registry used to publish it, and how it is distributed.
The first five rows list systems which cache bytecode compilation.
In Julia, type inference is the cached step, because resolving expression types to concrete type signatures is expensive.
In the JVM, Linking is the most important step to cache.
Android's ART runtime takes a more comprehensive approach by caching both  class initialization and native compilation together.
ART does so at two levels.
At the OS level, a single boot image covers all Android framework classes and is distributed as part of the OS image by device manufacturers.
It is shared across all applications on the device~\cite{jonathanlevin_artofdalvik_2016}.
At the application level, ART produces a local cache at install time that covers the application and all its dependencies together in a single monolithic observation run~\cite{noauthor_configure_2026}.
Although this application-level cache is stored locally on the device and not distributed, ART compiles it from a cloud profile aggregated from anonymized user telemetry, later merged with a profile recorded locally on the device~\cite{noauthor_configure_2026}.
None of the surveyed systems distributes execution caches through a package registry.
In this paper, we improve the state of the art of execution caching by introducing the first system to do so.

\begin{table}[t]
\centering
\caption{Execution caches in popular language runtimes.}
\label{tab:execution-caches}
\resizebox{\columnwidth}{!}{%
\begin{tabular}{@{}l l l l l@{}}
\toprule
\textbf{Language} & \textbf{Runtime} & \textbf{Cache} & \makecell[l]{\textbf{Stage}\\\textbf{Cached}} & \makecell[l]{\textbf{Registry}\\\textbf{Distribution}} \\
\midrule
Python     & CPython        & Pycache~\cite{barrywarsaw_pep_2009}                          & \makecell[l]{Bytecode\\Comp.}      & Per source file~\textcolor{red}{\ding{55}} \\
PHP        & Zend Engine    & OPcache~\cite{php_php_2026}                                  & \makecell[l]{Bytecode\\Comp.}      & Local~\textcolor{red}{\ding{55}} \\
Ruby       & CRuby          & bootsnap~\cite{rubyonrails_rails_2026}                       & \makecell[l]{Bytecode\\Comp.}      & Local~\textcolor{red}{\ding{55}} \\
JavaScript & V8             & V8 code cache~\cite{leszekswirski_code_2020}                 & \makecell[l]{Bytecode\\Comp.}      & Local~\textcolor{red}{\ding{55}} \\
Brainfuck  & BF Interp.     & Precompile cache~\cite{sunjayvarma_precompile_2017}          & \makecell[l]{Bytecode\\Comp.}      & Local~\textcolor{red}{\ding{55}} \\
Julia      & Julia          & \makecell[l]{Julia Precomp.\\Cache~\cite{marcellhavlik_julia_2021}} & \makecell[l]{Type\\Inference}      & Local~\textcolor{red}{\ding{55}} \\
Java       & \ac{JVM}       & AOT Cache~\cite{lam_jep_2025a}                               & Linking                            & Local~\textcolor{red}{\ding{55}} \\
Android    & ART            & Boot image~\cite{jonathanlevin_artofdalvik_2016}              & \makecell[l]{Native\\Comp.}        & OS Image~\textcolor{red}{\ding{55}} \\
Android    & ART            & \makecell[l]{App image~\cite{noauthor_configure_2026}}        & \makecell[l]{Native\\Comp.}        & Local~\textcolor{red}{\ding{55}} \\
Java       & \ac{JVM}       & \toolname{} (this work)                                      & Linking                            & Maven Central~\textcolor{green!60!black}{\ding{51}} \\
\bottomrule
\end{tabular}}
\end{table}

\begin{figure*}[t]
  \centering
  \includegraphics[width=\textwidth]{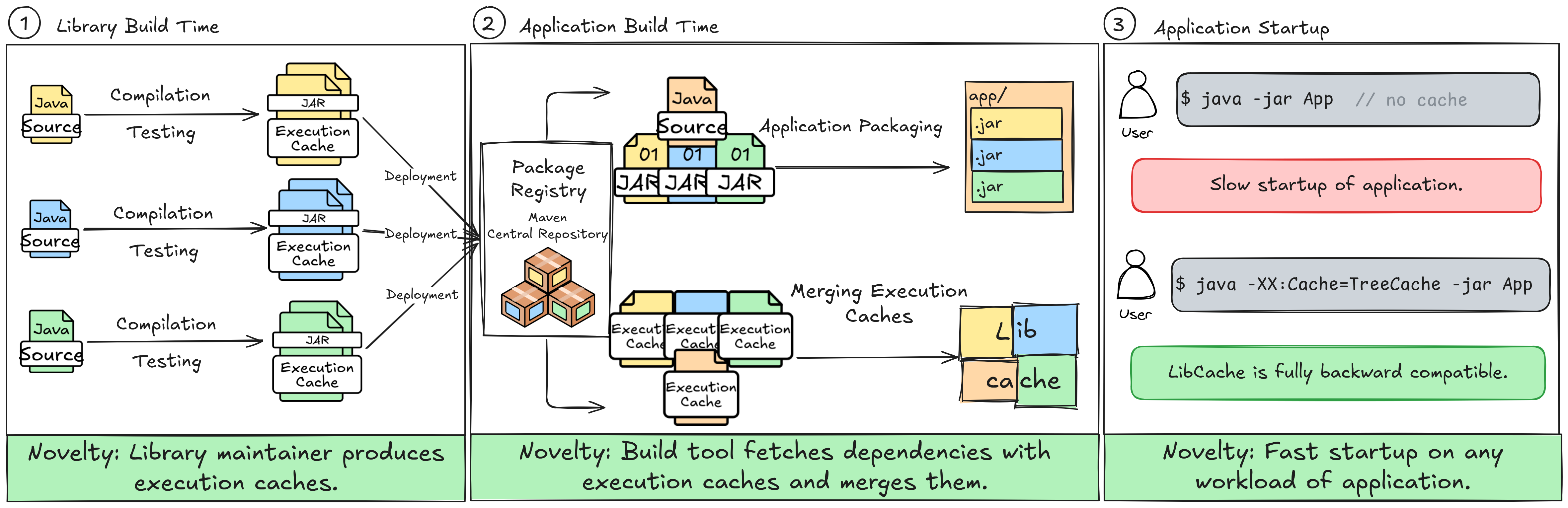}
  \caption{Overview of the \toolname{} workflow.}
  \label{fig:workflow}
\end{figure*}

\section{Problem Statement \& Motivating Example}
\label{sec:problem}

An execution cache is the artifact that a runtime produces during an observation run, serialized on disk for later reuse.
It captures some aspect of the work that the runtime performs before executing code, such as producing bytecode, resolving metadata, or compiling to native code, depending on the cached stage (\autoref{tab:execution-caches}).
At the next startup, the runtime loads the cache instead of redoing the preparation work, which makes startup faster.

Existing systems produce execution caches through a single, monolithic observation run~\cite{lam_jep_2025a}, which creates a fundamental problem.
The cache only covers the code exercised during that particular run and \textit{cannot generalize to other workloads}.
This is the problem we address in this paper.

Consider \texttt{commons-configuration2}, a library that reads configuration from properties files and XML files.
A developer generates an execution cache by running the library to read key-value pairs from a \texttt{.properties} file.
This cache does not contain the XML parsing classes, so an application that reads an XML configuration file gains no startup benefit from it.
This coverage problem has also been reported independently on the OpenJDK Leyden mailing list~\cite{ryannett_external_2026} and in JDK's own design notes~\cite{danheidinga_thoughts_2024}.
\toolname{} addresses this problem by producing one execution cache per dependency, using the dependency's own test suite as the workload, and merging all per-dependency caches at application build time.
For \texttt{commons-configuration2}, its library's test suite exercises both key-value parsing and XML configuration parsing, so the resulting execution cache covers both behaviors and benefits any downstream application workload regardless of which configuration format it uses.

\section{\toolname}

\subsection{Overview}
\autoref{fig:workflow} illustrates the \toolname{} workflow across its three main phases:
library maintainers produce per-library execution caches (\autoref{sec:per-library-cache}), application developers merge them at build time (\autoref{sec:application-building}), and the application consumes the \toolname{} cache at runtime startup (\autoref{sec:application-startup}).

\subsubsection{Per-library Execution Cache}
\label{sec:per-library-cache}
In the first phase, each library maintainer runs the library under its own test suite and produces an execution cache that captures the classes exercised during the test run.
This works well for well tested libraries where the tests are representative of real-world use~\cite{tiwari_production_2022}.
When a library ships no test suite, a minimal custom workload that exercises the library's public API serves as an alternative recording source.
In a \toolname{} workflow, the library maintainer then publishes the execution cache to a package registry such as Maven Central alongside the library JAR, making it available to all downstream consumers.

\subsubsection{Application Building}
\label{sec:application-building}
In the second phase, the application build proceeds in two parts (middle of \autoref{fig:workflow}).
The package manager downloads each dependency together with its distributed execution cache, realizing the per-dependency distribution model shown in \autoref{tab:execution-caches}.
The build tool then runs the application's own test suite and produces an execution cache that captures the classes exercised by the application.
The \toolname{} merge step (see \autoref{sec:treecache-impl}) then combines the execution caches of all dependencies with the application's own execution cache into a single \toolname{} cache.

\subsubsection{Application Startup}
\label{sec:application-startup}
In the third phase of the \toolname{} lifecycle, the application starts in production using the \toolname{} cache (right-hand side of `Application Build Time' in \autoref{fig:workflow}).
The \toolname{} cache contains all the metadata about the application and its dependencies that the runtime needs to execute.
This removes the initialization work that the runtime otherwise performs at startup in the absence of a cache, resulting in significant speedups.

\subsubsection{Recapitulation}  \toolname{} addresses the coverage problem stated in \autoref{sec:problem} through two complementary mechanisms.
The first mechanism, per-dependency cache production, lets each library developer record execution state against the library's own test suite, which broadens coverage beyond any single application workload.
The second mechanism, cache merging, creates a powerful aggregated cache that the runtime later consumes at startup.
The original lifecycle still works when dependencies evolve, as discussed in \autoref{sec:cache-lifecycle}.

\subsection{Merging Execution Caches}
\label{sec:merge-design}

Per-dependency cache distribution results in one execution cache per package.
This requires a merge step, which is one of the core novelties of this paper.

Recall that most runtimes in \autoref{tab:execution-caches} serialize each cache as a self-contained structured artifact.
For example, CPython's \texttt{.pyc} format stores a marshalled code object~\cite{barrywarsaw_pep_2009}, V8 serializes per-script bytecode blobs~\cite{leszekswirski_code_2020}, and Ruby's bootsnap dumps per-file YARV instruction sequences~\cite{rubyonrails_rails_2026}.
In each case, cross-module references are symbolic names that the respective module system resolves at load time.
So, regardless of the execution cache stack, per-dependency caches are usable only after a merge step.

The general merge algorithm for a memory-mapping runtime proceeds in three steps regardless of the specific runtime.
The first step is extraction: each library cache records portable symbolic identifiers that are independent of any memory layout; the merge collects these identifiers across all library caches to form a union symbol set.
The second step is reconstruction: the runtime processes the union symbol set to produce a new coherent memory image.
The third step is derivation: the merge regenerates all layout-dependent metadata from scratch against the merged layout.
\autoref{sec:treecache-impl} describes how \toolname{} implements these three steps.

\subsection{Compatibility with Dependency Updates}
\label{sec:cache-lifecycle}
Execution caches follow the same release lifecycle as the packages they accompany.
Each dependency owns its cache, so an update to one dependency replaces only that dependency's cache entry and leaves the caches of all other dependencies valid.

When a maintainer releases a new version of a library, they re-run the library's test suite and publish the resulting cache alongside the new package version.
The registry hosts the cache as an additional artifact under the same coordinates as the package, so a registry such as Maven Central supports cache distribution without structural changes.
At build time, when a library is upgraded, the package manager downloads both the new version of the code and the updated cache.
The build tool then re-runs the merge step, substituting the new cache entry while every other dependency's cache remains intact.
The change impact of a dependency update stays bounded by the dependency that changed rather than the full dependency graph.

\subsection{Implementation Details}

We implement \toolname{} as a fork of the \ac{JVM}'s existing AOTCache mechanism~\cite{lam_jep_2025a}.
We build our custom JDK on top of JDK 24.0.2.
It adds a merge mode that combines multiple AOTCaches into one, reusing the existing execution cache production.
This merge mode is a complex change of ~1,100 C++ LOC (1,222 added, 138 removed) spread over 38 files of the JDK.
Our implementation is open source and available in our fork of JDK\footnote{\url{https://github.com/algomaster99/jdk/blob/47a75e5b759c01fe03a9ef404ef59175d768ab5f/implementation.diff}}.

\begin{figure}[t]
\centering
\includegraphics[width=\columnwidth]{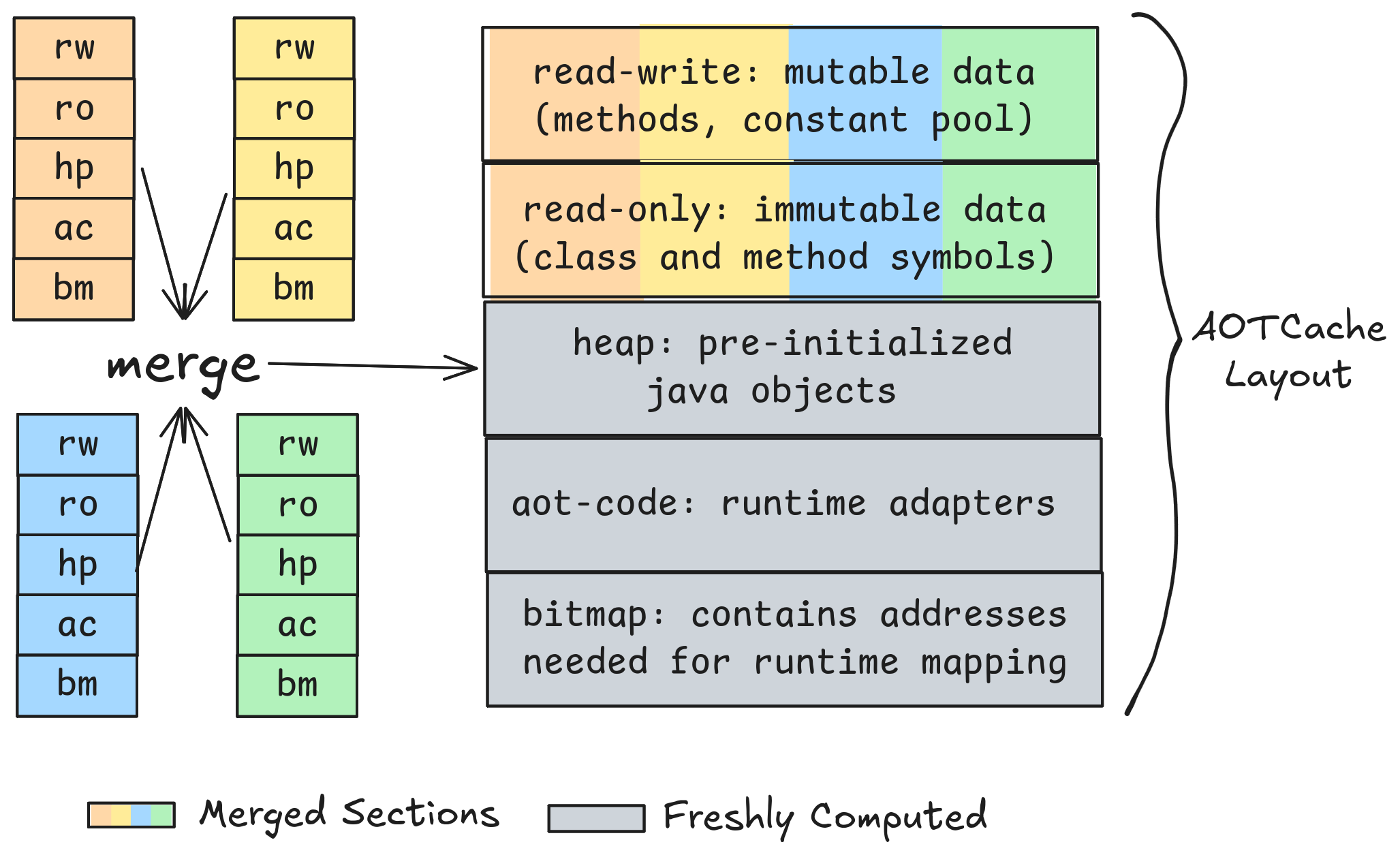}
\caption{The merge combines the RW and RO regions from multiple input caches into a single \toolname{} cache.
The HP, AC, and BM regions are regenerated from the merged class set rather than carried over from the inputs.}
\label{fig:aotcache-format}
\end{figure}

\subsubsection{AOTCache Data Structure}
\label{sec:aot-cache-structure}

The AOTCache data structure is introduced in JEP~483~\cite{lam_jep_2025a}.
Its goal is to reduce Java application startup time by making classes available in a linked state at \ac{JVM} startup.

To produce the cache, the user runs the application once with the \texttt{-XX:AOTCacheOutput} flag.
The \ac{JVM} observes all classes that are loaded and linked during this run and serializes their internal metadata into a memory-mapped file called the AOTCache file.
To consume the cache, the user runs the application with the \texttt{-XX:AOTCache} flag.

The AOTCache file contains five regions, illustrated in \autoref{fig:aotcache-format}.
The \textbf{Read-Write (RW)} region stores the mutable internal \ac{JVM} metadata structures such as methods and constant pool for each cached class.
The \textbf{Read-Only (RO)} region stores immutable metadata such as class names.
Together, the RW and RO regions let the \ac{JVM} skip loading and linking at startup.
Instead of searching the classpath, parsing bytecode, allocating these structures, verifying bytecode, and resolving constant-pool references, the \ac{JVM} memory-maps the pre-built structures.
The third region in the file is the \textbf{Heap (HP)} region with pre-initialized Java objects.
The fourth region is the \textbf{AOT Code (AC)} region, which stores pre-generated runtime adapters that let interpreted code invoke JIT-compiled code and vice versa.
Finally, the \textbf{Bitmap (BM)} region contains pointer relocation maps used to patch the other regions to their correct runtime addresses at load time and does not reduce startup time directly~\cite{baker_java_2025}.

\begin{algorithm}
\caption{Merging multiple AOTCaches into a single unified cache.}
\label{alg:merge}
\begin{algorithmic}[1]
\Input set of execution caches $\{C_1, \ldots, C_n\}$; application classpath
\Output \toolname{} cache file written to disk
\State $merged \gets \{\}$
\For{each cache $C_i$ in $\{C_1, \ldots, C_n\}$}
  \For{each class name $n$ recorded in $C_i$}
    \If{$n$ is resolvable on classpath \textbf{and} $n \notin merged$}
      \State $merged \gets merged \cup \{\mathrm{load\text{-}and\text{-}link}(n)\}$
    \EndIf
  \EndFor
\EndFor
\State serialize $merged$ into RW and RO regions
\State fork child \ac{JVM} to assemble final \toolname{} cache
\State regenerate HP from initialized Java objects in $merged$
\State regenerate AC adapters for all method signatures in $merged$
\State regenerate BM from memory layout of RW, RO, and HP
\State write final \toolname{} cache file
\end{algorithmic}
\end{algorithm}

\subsubsection{Merging Implementation}
\label{sec:treecache-impl}

The merge component of \toolname{} accepts a set of AOTCache files as input.
For each input cache, \toolname{} loads it, enumerates all archived class names, and writes them to a temporary file.
Next, \toolname{} reads all temporary files and, for each class name, locates the corresponding bytecode on the classpath, processes it through loading and linking, and adds the result to the merged set.
Class names not resolvable on the classpath are dropped, which accommodate hidden classes with runtime-generated names and test-harness classes absent from the application classpath. In practice, this means that those classes are never cached.

Lambda proxy classes need special care because their runtime-generated names are incompatible with name-based enumeration.
\toolname{} handles this by loading one input cache as the live archive at \ac{JVM} startup via memory mapping~\cite{kerrisk_linux_2018}.
The \ac{JVM} supports at most one live archive, so only this cache's lambda proxy classes become present as live objects in memory.
The archive builder discovers and includes them directly, without requiring name-based lookup.
As a result, \toolname{} is able to recover lambda proxy classes as part of the merged set.

The merged class set is serialized into the RW and RO regions of the AOTCache data structure (\autoref{fig:aotcache-format}).
Next, \toolname{} reconstructs the remaining regions that depend on RW and RO.
The HP region is regenerated because heap objects in the input caches embed pointers into their respective RW and RO layouts, which become invalid after merging; all pointers are recomputed.
Each runtime adapter, a stub that lets interpreted code invoke JIT-compiled code (\autoref{sec:aot-cache-structure}), is keyed to a specific method type signature.
The AC region is regenerated because the merged class set introduces type signatures that span the input caches, requiring adapters that no individual input cache covers.
The BM region is regenerated because it encodes the memory layout of the other regions.

\section{Experiments}

\subsection{Research Questions}
\autoref{fig:workflow} structures our evaluation around the three phases of the \toolname{} workflow.
The first phase (library build time) motivates RQ1, which measures the overhead that cache production adds to the library build.
The second phase (application build time) motivates RQ2, which covers both the storage cost of distributing caches through a package registry and the time cost of merging them at application build time.
The third phase (application startup) motivates RQ3, which evaluates whether the merged cache achieves fast startup across diverse workloads.
We answer the following research questions.

\begin{itemize}
  \item[\textbf{RQ1}] What build time overhead does \toolname{} cache production add to the build process?
  \item[\textbf{RQ2}] What is the overhead of the \toolname{} distribution model in terms of storage cost on the package registry and merge time at application build time?
  \item[\textbf{RQ3}] Does the \toolname{} cache achieve fast startup across diverse workloads?
\end{itemize}

\subsection{Dataset}

\begin{table}[t]
\centering
\caption{Dataset of Java projects used for evaluation.}
\label{tab:dataset}
\scriptsize
\begin{tabular}{@{}p{0.22\columnwidth} p{0.15\columnwidth} c p{0.38\columnwidth}@{}}
\toprule
\textbf{Project} & \textbf{Version} & \textbf{\# Deps} & \textbf{Workloads} \\
\midrule
batik                 & \href{https://central.sonatype.com/artifact/org.apache.xmlgraphics/batik-all/1.19}{1.19}  & 5 & svg-parse, svg-to-png, svg-to-jpeg, svg-generate \\
thymeleaf             & \href{https://central.sonatype.com/artifact/org.thymeleaf/thymeleaf/3.1.5.RELEASE}{3.1.5.RELEASE} & 4 & html-render, text-render, xml-render, fragment-render \\
biojava               & \href{https://central.sonatype.com/artifact/org.biojava/biojava-core/7.2.5}{7.2.5} & 6 & genbank-write, msa, aa-prop, pdb-parse \\
\bottomrule
\end{tabular}
\end{table}

We select projects according to three criteria.
First, the project must use Maven as its build system, as our prototype is implemented as a Maven plugin.
Second, the project's test suite must not use bytecode instrumentation, as instrumentation-based test frameworks are incompatible with our cache~\cite{ioilam_jdk8387190_2026} (see \autoref{sec:threats}).
Third, the project must expose multiple primary use cases documented by the library authors, so the evaluation covers meaningfully distinct workloads.
These constraints reduce the candidate pool substantially; we discuss their implications for generalizability in \autoref{sec:threats}.
For each selected project, we select one workload per primary documented use case of the library (e.g., for \textsc{BioJava}, the alignment, structure, and sequence-I/O modules are each represented by one workload~\cite{lafita_biojava_2019}).

As a result of the above criteria, we collect three real-world Java projects, shown in \autoref{tab:dataset}.
For \texttt{batik}, \texttt{svg-parse} loads an SVG file into a DOM tree, \texttt{svg-to-png} and \texttt{svg-to-jpeg} rasterise it to image formats, and \texttt{svg-generate} produces an SVG from programmatic drawing operations.
For \texttt{thymeleaf}, each workload renders a template in a different output mode, covering HTML, plain text, XML, and fragment inclusion.
For \texttt{biojava}, \texttt{genbank-write} serialises a parsed sequence record into two standard bioinformatics file formats, \texttt{msa} aligns multiple biological sequences, \texttt{aa-prop} computes physicochemical properties of a sequence from an XML-defined lookup table, and \texttt{pdb-parse} parses a structural biology file to extract residue and atom data.

\section{RQ1: Build Time Overhead of \toolname{} Cache Production}

\subsection{Objective}

The goal of this research question is to quantify the additional build time that \toolname{} cache production adds to the standard build process.
We measure the percentage increase in total build time when cache production is enabled, compared to a baseline build that produces no cache.
We assess whether this overhead is small enough to make cache production practical for routine library releases.

\subsection{Methodology}

We use two approaches to measure per-dependency cache production overhead, depending on whether the dependency ships a test suite.
We measure the wall-clock time of each approach and report absolute times and the percentage overhead relative to the baseline.

\textbf{Test suite.}
For each dependency that ships a test suite, we measure the baseline as the wall-clock time of a standard Maven testing run (\texttt{mvn clean test}).
We then enable cache production, which records execution state into a per-dependency cache file during the test run, and measure the wall-clock time of this augmented run.
The overhead is the difference between the two runs.

\textbf{Custom workload.}
Five dependencies in our dataset ship no test suite (as described in \autoref{sec:per-library-cache}): \texttt{commons-logging}, \texttt{xml-apis}, and \texttt{xml-apis-ext} for \emph{batik}; \texttt{unbescape} for \emph{thymeleaf}; and \texttt{forester} for \emph{biojava}.
For each, we write a minimal custom workload that exercises the library's public API.
The baseline is a plain \texttt{java -jar} run; the cache production run adds cache recording to the same JVM invocation.
These entries are marked $^*$ in \autoref{tab:rq1-overhead}.

\subsection{Results}

\autoref{tab:rq1-overhead} reports the build time overhead for each execution cache produced in our dataset.

\textbf{Per-module overhead.}
\toolname{} adds between +10\% and +47\% overhead to the build time of the main application module in each project.
Dependencies without a test suite show inflated percentage overheads, but their absolute baseline times are under 0.2\,s, so each contributes at most 0.8\,s of additional build time, which is arguably acceptable in a typical multiminute Java build.

\subsection{Answer to RQ1}

\begin{mdframed}[backgroundcolor=gray!10, linewidth=1pt]
\textbf{Answer to RQ1:} \toolname{} cache production adds between +10\% and +47\% to the build time of the main application module; dependencies vary more widely depending on test suite size and presence. It is a reasonable price to pay to enter a world where package distribution encompasses production of code and execution caches.
\end{mdframed}

\begin{table}[t]
\centering
\caption{Build time overhead of \toolname{} cache production per module across all study subjects.
$^*$~Cache is produced from a custom workload because this dependency has no test suite.}
\label{tab:rq1-overhead}
\resizebox{\columnwidth}{!}{%
\begin{tabular}{@{}l r r r@{}}
\toprule
\textbf{Module} & \textbf{Baseline (s)} & \textbf{Cache prod. (s)} & \textbf{Overhead (\%)} \\
\midrule
\multicolumn{4}{@{}l}{\textit{batik}} \\
batik               & 20.5 & 29.6 & +44.3 \\
xmlgraphics-commons & 11.1 & 19.8 & +78.5 \\
commons-io          & 82.0 & 92.8 & +13.2 \\
commons-logging     &  0.1$^*$ &  0.9 & +800.0 \\
xml-apis            &  0.1$^*$ &  0.9 & +607.1 \\
xml-apis-ext        &  0.1$^*$ &  0.6 & +850.7 \\
\midrule
\multicolumn{4}{@{}l}{\textit{thymeleaf}} \\
thymeleaf           & 121.7 & 134.4 & +10.4 \\
attoparser          &  61.1 &  64.8 &  +6.0 \\
ognl                &  14.1 &  19.6 & +39.1 \\
slf4j-api           &   7.2 &   9.3 & +27.9 \\
unbescape           &   0.1$^*$ &   0.7 & +430.1 \\
\midrule
\multicolumn{4}{@{}l}{\textit{biojava}} \\
biojava      &  51.9 &  76.2 & +46.8 \\
biojava-alignment &  11.2 &  25.8 & +129.4 \\
biojava-aa-prop   &   9.5 &  27.1 & +187.0 \\
biojava-structure & 111.4 & 134.4 & +20.7 \\
commons-codec     &   9.1 &  16.5 & +82.0 \\
slf4j-api         &  12.0 &  14.6 & +21.6 \\
forester          &   0.1$^*$ &   0.8 & +881.2 \\
\bottomrule
\end{tabular}%
}
\end{table}

\section{RQ2: Distribution Overhead of \toolname{}}
\label{sec:rq2}

\subsection{Objective}

The goal of this experiment is to characterize the overhead of the \toolname{} distribution model along two dimensions: the storage cost of hosting caches on a package registry, and the time cost of merging them at application build time.
We assess whether both costs are bounded enough to make the model practical.

\subsection{Methodology}

As described in \autoref{fig:workflow}, in \toolname{}, each dependency is distributed as a bundle of code and execution data.
This necessarily has a storage impact.
We analyze 21 execution caches: 3 application caches, 15 dependency caches, and 3 \toolname{} merged caches across the three projects in the dataset (\autoref{tab:dataset}).

For each dependency cache, we measure its size on disk.
We compare this against the size of the dependency's JAR file as published on Maven Central, which represents the binary code shipped with that dependency.

Each AOT Cache can be pretty-printed to a human-readable dump of the cache contents that the JVM produces as a built-in diagnostic.
We parse these map files to count the number of archived \texttt{Class} objects and to classify every recorded method as either a JDK method (packages \texttt{java.*}, \texttt{jdk.*}, \texttt{sun.*}, \texttt{javax.*}, \texttt{com.sun.*}) or an application class, reporting the fraction of library methods as App\%.
To extract the class name from a method signature accurately, we strip the return type prefix and split on the last dot before the opening parenthesis.

Finally, we record the wall-clock duration of the \toolname{} merge step for each project; because the merge step has no baseline equivalent, we report only its absolute time.

\subsection{Results}

\autoref{tab:rq2-sizes} reports cache sizes and class-count breakdowns across all 21 analyzed caches (18 real caches plus the \toolname{} merged cache for each project).

\textbf{Cache size versus code size.}
Across the 18 analyzed libraries, execution caches are substantially larger than the corresponding JAR files.
The overhead is at least 3$\times$ even for the largest library in our dataset (\texttt{biojava}: 12\,MB JAR, 40\,MB cache), and exceeds 300$\times$ for small wrapper libraries (\texttt{biojava-aa-prop}: 0.1\,MB JAR, 32\,MB cache).
On average across our dataset, an execution cache is 18$\times$ the size of the corresponding JAR, suggesting that hosting \toolname{} caches on Maven Central would multiply its JAR storage footprint by a similar factor.

\textbf{Merged cache versus App cache.}
The \toolname{} merged cache contains fewer application classes than the App cache in most projects (\emph{batik}: 1,964 vs.\ 2,085; \emph{thymeleaf}: 1,231 vs.\ 4,012).
The merge algorithm drops any class not resolvable on the application classpath (Algorithm~\ref{alg:merge}, line~3), excluding test-harness classes (e.g., Surefire) that each dependency's test suite loads but that are absent from the production classpath.
Additionally, each class is included at most once in the merged cache even when multiple dependency caches archive it, so cross-dependency duplicates do not inflate the count. This is why the \toolname{} merged cache contains fewer application classes than the App cache alone for \emph{batik} and \emph{thymeleaf}.

\textbf{JDK classes dominate every cache.}
In libraries such as \texttt{commons-logging} and \texttt{slf4j-api}, the JDK contributes 1,375 and 1,552 classes respectively against only 18 and 63 application classes.
Even in larger dependencies such as \texttt{commons-io}, JDK classes (4,645) outnumber application classes (1,354) by more than three to one.
This dominance compounds across dependency caches, since every test suite triggers overlapping JDK subsystems and each cache independently re-archives the same JDK classes.
For \emph{batik}, the five dependency caches together archive 11,640 JDK classes yet the \toolname{} merged cache holds only 5,734.
For \emph{thymeleaf}, 6,318 classes across four caches reduce to 2,793.
For \emph{biojava}, 14,365 classes across six caches reduce to 5,680.
Across all three projects, 51--60\% of the JDK classes archived across dependency caches are duplicates.
A single shared JDK cache would eliminate this redundancy and represent a clear optimization opportunity for the \toolname{} distribution model.

\textbf{Time taken by merge step.}
Beyond storage, the merge step adds 20.9\,s for \emph{batik}, 16.0\,s for \emph{thymeleaf}, and 23.2\,s for \emph{biojava}; its cost scales with the total unique class count across all input caches.

\begin{table}[t]
\centering
\caption{Cache size and archived class count broken down by JDK and application classes.
Each group header names the project as listed in the dataset (Table~\ref{tab:dataset}).
The first row of each group is the cache built from the main application's test suite.
The remaining rows are the caches of its dependencies.
\toolname is our merged cache.}
\label{tab:rq2-sizes}
\begin{tabular}{@{}l r r r r@{}}
\toprule
\textbf{Module} & \multicolumn{2}{c}{\textbf{Size}} & \multicolumn{2}{c}{\textbf{Classes}} \\
\cmidrule(lr){2-3}\cmidrule(l){4-5}
 & \textbf{Code} & \textbf{Execution} & \textbf{JDK} & \textbf{App} \\
\midrule
\multicolumn{5}{@{}l}{\textit{batik}} \\
App: batik               & 5.7\,MB & 41\,MB &  3,567 &  2,085 \\
Dep: xmlgraphics-commons & 0.7\,MB & 28\,MB &  3,180 &    645 \\
Dep: commons-io          & 0.5\,MB & 39\,MB &  4,645 &  1,354 \\
Dep: commons-logging     & 0.1\,MB & 13\,MB &  1,375 &     18 \\
Dep: xml-apis            & 0.2\,MB & 13\,MB &  1,442 &     34 \\
Dep: xml-apis-ext        & 0.1\,MB & 10\,MB &    998 &     10 \\
\toolname   batik        & --- & \textbf{54\,MB} &  5,734 &  1,964 \\
\midrule
\multicolumn{5}{@{}l}{\textit{thymeleaf}} \\
App: thymeleaf           & 2.4\,MB & 43\,MB &  2,721 &  4,012 \\
Dep: attoparser          & 0.2\,MB & 14\,MB &  1,533 &    168 \\
Dep: ognl                & 0.3\,MB & 21\,MB &  2,241 &    320 \\
Dep: slf4j-api           & 0.1\,MB & 14\,MB &  1,552 &     63 \\
Dep: unbescape           & 0.2\,MB & 10\,MB &    992 &     83 \\
\toolname thymeleaf      & --- & \textbf{29\,MB} &  2,793 &  1,231 \\
\midrule
\multicolumn{5}{@{}l}{\textit{biojava}} \\
App: biojava        & 12\,MB & 40\,MB &  4,653 &  1,147 \\
Dep: biojava-alignment   & 0.1\,MB & 25\,MB &  2,333 &    945 \\
Dep: biojava-aa-prop     & 0.1\,MB & 32\,MB &  2,518 &  1,696 \\
Dep: biojava-structure   & 1.4\,MB & 55\,MB &  4,779 &  3,527 \\
Dep: forester            & 0.1\,MB & 12\,MB &  1,180 &     16 \\
Dep: commons-codec       & 0.1\,MB & 17\,MB &  2,007 &     47 \\
Dep: slf4j-api           & 0.1\,MB & 14\,MB &  1,548 &     63 \\
\toolname   biojava & --- & \textbf{57\,MB} &  5,680 &  1,999 \\
\bottomrule
\end{tabular}
\end{table}

\subsection{Answer to RQ2}

\begin{mdframed}[backgroundcolor=gray!10, linewidth=1pt]
\textbf{Answer to RQ2:} The \toolname{} distribution model incurs overhead on two fronts: execution caches are on average 18$\times$ larger than the corresponding JAR files, dominated by JDK classes duplicated across 51--60\% of dependency caches, and the merge step takes 16--23\,s per project, scaling with the total unique class count rather than the number of dependencies; a layered cache design separating JDK from application classes would substantially reduce both.
\end{mdframed}

\begin{table*}[t]
\centering
\caption{Mean startup time (ms) $\pm$ standard deviation over 30 runs per workload. Speedup is relative to the default (no cache).}
\label{tab:rq3-timing}
\resizebox{\textwidth}{!}{%
\begin{tabular}{l l r r r r r}
\toprule
 & & \multicolumn{1}{c}{\textbf{No Cache}} & \multicolumn{2}{c}{\textbf{AOTCache}} & \multicolumn{2}{c}{\textbf{\toolname}} \\
\cmidrule(lr){3-3} \cmidrule(lr){4-5} \cmidrule(lr){6-7}
\textbf{Project} & \textbf{Workload} & \textbf{Time (ms)} & \textbf{Time (ms)} & \textbf{Speedup ($\times$)} & \textbf{Time (ms)} & \textbf{Speedup ($\times$)} \\
\midrule
\multirow{5}{*}{batik}
  & svg-parse & $186.8 \pm 8.4$ & $130.2 \pm 20.9$ & 1.43x & $128.6 \pm 3.3$ & \textbf{1.45x} \\
  & svg-to-png & $693.3 \pm 19.1$ & $577.6 \pm 88.2$ & 1.20x & $418.0 \pm 9.1$ & \textbf{1.66x} \\
  & svg-to-jpeg & $686.0 \pm 21.5$ & $568.1 \pm 95.0$ & 1.21x & $377.1 \pm 10.5$ & \textbf{1.82x} \\
  & svg-generate & $265.5 \pm 10.9$ & $215.7 \pm 32.9$ & 1.23x & $177.9 \pm 6.1$ & \textbf{1.49x} \\
  & \textit{Average} & & & 1.27x & & \textbf{1.60x} \\
\midrule
\multirow{5}{*}{thymeleaf}
  & html-render & $368.4 \pm 7.9$ & $235.2 \pm 19.2$ & 1.57x & $233.8 \pm 6.8$ & \textbf{1.58x} \\
  & text-render & $312.7 \pm 8.0$ & $185.8 \pm 22.0$ & 1.68x & $163.8 \pm 4.0$ & \textbf{1.91x} \\
  & xml-render & $302.2 \pm 9.2$ & $173.2 \pm 14.5$ & 1.74x & $164.8 \pm 7.0$ & \textbf{1.83x} \\
  & fragment-render & $405.8 \pm 10.5$ & $273.0 \pm 17.9$ & 1.49x & $245.5 \pm 6.0$ & \textbf{1.65x} \\
  & \textit{Average} & & & 1.62x & & \textbf{1.74x} \\
\midrule
\multirow{5}{*}{biojava}
  & genbank-write & $147.5 \pm 8.5$ & $130.6 \pm 10.2$ & \textbf{1.13x} & $141.8 \pm 5.1$ & 1.04x \\
  & msa & $177.9 \pm 5.1$ & $154.2 \pm 7.3$ & \textbf{1.15x} & $158.0 \pm 3.9$ & 1.13x \\
  & aa-prop & $318.8 \pm 5.9$ & $266.9 \pm 51.1$ & 1.19x & $216.4 \pm 6.0$ & \textbf{1.47x} \\
  & pdb-parse & $596.1 \pm 15.1$ & $552.3 \pm 44.7$ & 1.08x & $448.3 \pm 8.3$ & \textbf{1.33x} \\
  & \textit{Average} & & & 1.14x & & \textbf{1.24x} \\
\bottomrule
\end{tabular}%
}
\end{table*}

\section{RQ3: Performance Comparison Against AOTCache}

\subsection{Objective}

The goal of this research question is to evaluate the performance enhancements of \toolname{} compared to a simple AOTCache built from a single workload.

\subsection{Methodology}

For each study subject and workload listed in Table~\ref{tab:dataset}, we evaluate three configurations: the default \ac{JVM} with no cache, AOTCache, and \toolname{}.

To account for randomness in the measurement, we run each workload 30 times~\cite{arcuri_practical_2011}.
For the default configuration, we run each workload without any cache and report the mean startup time and standard deviation.
For AOTCache, we perform a cross-workload evaluation: each project has $K=4$ workloads, so for each workload~$w$, we run it with each of the $K-1=3$ caches produced from the other workloads for 30 runs each, and report the mean startup time pooled across all $30 \times (K-1)$ runs.
For \toolname{}, we produce a \toolname{} cache (see \autoref{sec:treecache-impl}). We then run each workload 30 times and report the mean startup time and standard deviation.

We compare the three configurations using mean startup time (in milliseconds) and standard deviation over 30 runs.
We compute speedup as the ratio of the no-cache mean startup time to the mean startup time under a given configuration, reported per workload and averaged across workloads within each project.
For AOTCache, the mean and standard deviation are pooled across all $30 \times (K-1)$ cross-workload runs, so the reported value reflects the expected performance when the cache is produced from an arbitrary other workload rather than the one being tested.
A speedup greater than $1\times$ indicates that the configuration reduces startup time relative to the no-cache baseline; a higher speedup reflects a greater reduction.
We highlight the higher speedup in each row to show which configuration wins per workload.

\begin{table}[t]
\centering
\caption{Class-load source breakdown for \texttt{biojava} workloads.
\emph{Archived} counts classes served from the shared-objects archive;
\emph{Classpath} counts classes loaded from jars or the JDK module system;
\emph{Gen.}\ counts runtime-generated classes (lambdas, proxies).
In the no-cache baseline the \emph{Archived} count reflects only JDK classes from the default CDS archive; all application classes load from the classpath.}
\label{tab:rq3-classload}
\scriptsize
\begin{tabular}{@{}ll rrr r@{}}
\toprule
\textbf{Workload} & \textbf{Scenario} & \textbf{Archived} & \textbf{Classpath} & \textbf{Gen.} & \textbf{Total} \\
\midrule
\texttt{genbank-write} & No cache & 1,072 & 125 & 14 & 1,211 \\
 & AOTCache & 1,293 & 0 & 1 & 1,294 \\
 & \toolname{} & \textbf{6,866} & \textbf{11} & \textbf{0} & \textbf{6,877} \\
\midrule
\texttt{msa} & No cache & 996 & 195 & 11 & 1,202 \\
 & AOTCache & 1,300 & 0 & 1 & 1,301 \\
 & \toolname{} & \textbf{6,866} & \textbf{11} & \textbf{2} & \textbf{6,879} \\
\midrule
\texttt{aa-prop} & No cache & 1,083 & 865 & 73 & 2,021 \\
 & AOTCache & 2,043 & 0 & 32 & 2,075 \\
 & \toolname{} & \textbf{6,866} & \textbf{11} & \textbf{37} & \textbf{6,914} \\
\midrule
\texttt{pdb-parse} & No cache & 1,218 & 1,105 & 172 & 2,495 \\
 & AOTCache & 2,539 & 10 & 42 & 2,591 \\
 & \toolname{} & \textbf{6,866} & \textbf{18} & \textbf{122} & \textbf{7,006} \\
\midrule
\end{tabular}
\end{table}

\subsection{Results}

\autoref{tab:rq3-timing} shows the results of this evaluation.
The table reports the mean startup time (ms) and standard deviation over 30 runs for each workload under the three configurations, with speedup relative to the default (no cache).
For example, for \texttt{batik} with workload \texttt{svg-to-jpeg}, \toolname{} achieves a 1.82$\times$ speedup over the default, compared to 1.21$\times$ for AOTCache.

\toolname{} achieves a higher average speedup than AOTCache across all three projects, with the largest gap in \texttt{batik} (1.60$\times$ vs.\ 1.27$\times$).
\toolname{} merges caches from each dependency's own test suite, covering a broader class set than any single workload can.
This broader coverage explains the larger gap for \texttt{batik}, whose dependencies have richer test suites.

AOTCache shows a numerically higher speedup on two individual \texttt{biojava} workloads: \texttt{genbank-write} (1.13$\times$ vs.\ 1.04$\times$) and \texttt{msa} (1.15$\times$ vs.\ 1.13$\times$), though the difference for \texttt{msa} falls within measurement noise given the overlapping standard deviations.
We report the class-load breakdown for \texttt{biojava} because it is the only project where individual workloads show an inversion.
For \texttt{batik} and \texttt{thymeleaf}, \toolname{} wins on all workloads and the timing data alone is sufficient.
\autoref{tab:rq3-classload} explains this inversion.
\toolname{} eagerly loads all 6,866 classes from its archive at JVM startup regardless of which workload runs.
For \texttt{genbank-write} and \texttt{msa}, the workload requires only 1,202--1,211 classes in total; loading more than 5$\times$ as many pre-linked classes as the workload needs imposes a startup overhead that outweighs the AOT-linking benefit.
The heavier \texttt{aa-prop} and \texttt{pdb-parse} workloads require a broader class set, so \toolname{}'s comprehensive archive earns back the overhead and surpasses AOTCache.

\begin{mdframed}[backgroundcolor=gray!10, linewidth=1pt]
\textbf{Answer to RQ3:} \toolname{} enables faster application startups than no cache and AOTCache. This is thanks to the comprehensive cache provided by the dependencies' runs.
\end{mdframed}

\section{Related Work}
\label{sec:related}

\subsection{Reducing Startup and Warmup in Java}

\subsubsection{Pre-computed Execution State}

OpenJDK's Class Data Sharing (CDS) archives pre-processed, pre-verified JDK bootstrap class metadata into a shared archive, allowing multiple JVM processes to map it as read-only memory~\cite{noauthor_class_2025}.
AppCDS (JEP~310)~\cite{lam_jep_2023} extends this to application and dependency classes by recording observed classes during a trial run and dumping them into an archive.
JEP~350~\cite{zhou_jep_2021} later automated this process by capturing classes at JVM exit instead~\cite{ionutbalosin_application_2022}.
Project Leyden extends this model to also archive the results of class linking, and captures ahead-of-time method profiles to accelerate JIT compilation from the first run~\cite{lam_jep_2025a,igorveresov_jep_2026}.
The AOTCache maintainers also prototype \emph{iterative training} to let a cache absorb more than one observation run~\cite{ioilam_jdk8353598_2026}, but this prototype still composes runs sequentially rather than merging independently-produced per-dependency caches.

Eclipse OpenJ9 similarly provides a persistent shared class cache that stores pre-verified bytecode shared across JVM processes on the same machine~\cite{bhattacharya_improving_2017}.
Libra~\cite{ammons_libra_2007} takes this further by statically linking J9 with a minimal library OS into a single hypervisor partition image, precomputing the entire execution environment rather than just class metadata.
Azul's ReadyNow, a proprietary feature of the Azul JVM, records profiling data and compilation decisions from one run and replays them in subsequent runs to eliminate cold-start recompilation~\cite{delporte_how_2025}.
Majo et al.~\cite{majo_integrating_2017} persist JIT profiles at JVM exit in OpenJDK's HotSpot and replay them on the next run to skip re-profiling, while Zhang et al.~\cite{zhang_serverless_2021} record touched methods and profile data via Java Flight Recorder in a staging environment and feed them into the \ac{AOT} compiler for profile-guided optimization.

Each of these approaches builds its archive from a single trial run, capturing only the state observed under that workload.
In \toolname{}, the caches are composable and are first-class artifacts distributed through package repositories.

Prior evaluations of these approaches commonly use the DaCapo benchmark suite~\cite{blackburn_rethinking_2025}, whose workloads stress throughput and memory behavior under sustained execution.
The biojava workloads, for example, execute through a single entry class (\texttt{PeptidePropertiesImpl}) across all three input sizes, so any cache produced from one size covers the full class set of every other.
These workloads do not expose cross-workload class-loading diversity, and our evaluation therefore uses real-world projects whose workloads invoke disjoint library subsystems.

\subsubsection{Offloaded JIT Compilation}

JIT compilation can be heavily memory-intensive because the compiler allocates large buffers for compilation threads and intermediate representations.
Khrabrov et al.~\cite{khrabrov_jitserver_2022} address this with JITServer, which moves JIT compilation to a dedicated remote service so that client JVM instances share compiled artifacts across a network while keeping their own memory footprint small.
Our proposed approach takes a different angle by eliminating the loading and linking phases for dependency classes entirely, reducing the number of JVM steps that execute at startup rather than offloading one step to a remote service.

\subsubsection{Process Snapshot Reuse}

Kawachiya et al.~\cite{kawachiya_cloneable_2007} propose the Cloneable JVM, which forks an already-initialized JVM process image so that new isolated applications start 4 to 170 times faster than cold starts.
CRaC (Coordinated Restore at Checkpoint)~\cite{openjdk_openjdk_2026,azul_what_2024} takes a similar approach within OpenJDK by providing APIs that allow an application to coordinate the checkpointing of its live state and restore it in subsequent launches.
Both approaches require a fully warmed, running application as the source.
This is not the case for our approach as inputs of \toolname{} are per-dependency caches that are produced independently and composed later on.

\subsubsection{Compile-to-Native Ahead-of-Time Compilation}

GraalVM Native Image performs closed-world static analysis at build time to compile the application and its dependencies into a standalone native executable, eliminating JVM startup and warmup entirely~\cite{wimmer_initialize_2019}.
However, Native Image compiles without runtime feedback, so it cannot apply the speculative profile-guided optimizations that HotSpot's JIT performs adaptively, leaving peak throughput lower for long-running workloads~\cite{wade_aot_2017}.
The proposed approach retains full compatibility with the JVM's JIT compiler, preserving the adaptive profile-guided optimizations that Native Image cannot apply.

\subsection{Execution Caching in the Android Ecosystem}

\subsubsection{Dalvik Virtual Machine}

The original Android runtime, the Dalvik Virtual Machine (DVM), uses a register-based bytecode format and a trace-based \ac{JIT} compiler~\cite{oh_bytecodetoc_2015}.
Several works proposed ahead-of-time alternatives for Dalvik to decrease runtime overhead.
Lim et al.\ propose a selective \ac{AOT} compiler that profiles hot methods and compiles them statically while routing the remaining code through the \ac{JIT}~\cite{lim_selective_2012}.
Oh et al.\ present a bytecode-to-C compilation scheme that translates hot Dalvik methods to C and compiles them alongside the VM source~\cite{oh_bytecodetoc_2015}.
Kim et al.\ design a static Dalvik bytecode optimization framework that uses LLVM to produce optimized bytecode~\cite{kim_static_2015}.
These works target Dalvik and narrowly focus on improving compilation performance for individual applications. They do not address caching and distribution of runtime profiles across the software supply chain.

\subsubsection{Android Runtime (ART)}

ART, which supersedes Dalvik, determines which methods to \ac{AOT}-compile into the app image (\autoref{sec:execution-caches}) through profile-guided compilation, using a cloud profile distributed by the Play Store and a local profile recorded on the device~\cite{noauthor_configure_2026,calinjuravle_improving_2019}.
Our approach, targeting the \ac{JVM} ecosystem, merges pre-linked class metadata caches produced independently per dependency, relying on application and dependency test suites to drive cache production rather than cloud profiles aggregated from user telemetry.

Two works extend or build on ART to improve its compilation and caching behavior.
ShareJIT~\cite{xu_sharejit_2018} extends ART to share a single global \ac{JIT} code cache across multiple applications and processes, significantly reducing \ac{JIT} compilation time and memory consumption at the cost of limiting per-context specialization.
JavART~\cite{wang_javart_2025} proposes a lightweight rule-based \ac{JIT} compiler for ART that extracts translation rules through a learning approach, enabling fast and predictable compilation without the overhead of traditional optimization pipelines.
Unlike these works, our approach stores platform-neutral class metadata rather than native machine code, and distributes per-dependency caches through standard package registries such as Maven Central.

\section{Discussion}

\subsection{Threats to Validity}
\label{sec:threats}

\paragraph{Internal validity.}
The test suites used to populate per-dependency caches may not faithfully represent production class-access patterns, so measured startup improvements may not generalize to workloads that exercise subsystems the tests do not cover.

\paragraph{External validity.}
The primary threat to external validity is the small dataset.
Candidate selection is constrained because the build system must match our prototype's implementation restrictions, including custom class-loading and incompatibility with instrumentation-based test dependencies such as Mockito~\cite{ioilam_jdk8387190_2026}.

\subsection{Cache integrity in Supply Chain}
Distributing AOT caches through package registries introduces a supply-chain integrity concern.
Ohm et al.~\cite{ohm_constrictor_2026} show that attackers can inject malicious bytecode into \texttt{.pyc} files distributed on PyPI without failing the interpreter's own integrity checks.
Reproducible builds research shows that binary artifacts distributed through registries must be verifiable against their source~\cite{sharma_causes_2025}, and execution caches face the same requirement.
A tampered AOT cache could pre-load classes whose bytecode differs from the corresponding JAR.
An open direction is whether the JVM can verify at load time that each cached class matches its on-disk bytecode, preserving the supply-chain trust model for caches distributed through Maven Central.

\subsection{Dependency Selection for Cache Production}
RQ1 shows that cache production overhead varies across dependencies, with the highest relative values appearing for dependencies that lack a test suite and require a custom workload (\autoref{tab:rq1-overhead}).
RQ2 reveals that several of these same dependencies contribute relatively few unique application classes to the merged cache: \texttt{xml-apis-ext} contributes only 10 application classes, \texttt{commons-logging} only 18, and \texttt{forester} only 16 (\autoref{tab:rq2-sizes}).
A practical selection criterion is to omit a dependency from cache production when its unique application class count falls below a non-negligible threshold, forgoing only a marginal benefit to the merged cache.
Automating this selection based on per-dependency class counts is a promising direction for further reducing build overhead.

\subsection{Layered Execution Caches}
\label{sec:layered-caches}
The current design produces a single flat merged cache.
A layered structure would separate JDK classes, dependency classes, and application classes into independent layers that the JVM stacks at startup, allowing each layer to be published and consumed independently through a package registry.
Since JDK classes dominate cache size (\autoref{sec:rq2}), a shared JDK layer would eliminate the redundancy of embedding them in every application's cache.
The iterative training prototype in the Leyden project~\cite{ioilam_jdk8353598_2026} takes a step in this direction by letting a cache absorb multiple observation runs sequentially, but it composes runs one after another rather than maintaining independent layers that can be replaced or shared individually.

\section{Conclusion}

In this paper, we introduce \toolname{}, which addresses the limited coverage of existing
execution caches by distributing per-dependency caches through Maven Central and merging
them at build time into a single unified cache.
Across three real-world Java projects and twelve workloads, \toolname{} achieves up to 1.91$\times$ startup speedup and outperforms a cache built from a single workload in 10 of 12 workloads.
Cache production adds 10--47\% to the build time of the main application module, with dependencies varying by test suite size and presence.
Execution caches average 18$\times$ the JAR size.
Future work includes a layered cache design to eliminate JDK class duplication across
dependency caches, selective dependency inclusion to reduce build overhead, and JVM-side
bytecode verification to preserve the supply chain trust model of Maven Central.

\bibliographystyle{IEEEtran}
\bibliography{main}

\end{document}